\newcommand*{\wn}{cm$^{-1}$}
\newcommand*{\Hm}{H$_{2}$}
\newcommand*{\Dm}{D$_{2}$}
\newcommand*{\Hmi}{H$_{2}^{+}$}
\newcommand*{\HDi}{HD$^{+}$}
\newcommand*{\et}{\emph{et al.}}

\def\apj{Astroph.\ J.\ }

\def\apb{Appl.\  Phys.\  B }

\def\pla{Phys.\  Lett.\ A }

\def\jpb{J. Phys.\ B }

\def\rmp{Rev.\ Mod.\ Phys. }

\def\cjp{Can.\ J. Phys.\ }
\def\pccp{Phys.\ Chem.\ Phys.\ Chem.\ }

\documentclass[aps,prd,twocolumn,superscriptaddress,showpacs]{revtex4}
\usepackage{graphicx}
\usepackage{braket}
\usepackage{xfrac}
\usepackage{verbatim}
\usepackage{color,soul}

\begin{document}

\title{
Bounds on fifth forces from precision measurements on molecules}

\author{E. J. Salumbides}
\affiliation{Department of Physics and Astronomy, and LaserLaB, VU University, De Boelelaan 1081, 1081 HV Amsterdam, The Netherlands}
\affiliation{Department of Physics, University of San Carlos, Cebu City 6000, Philippines}
\author{J. C. J. Koelemeij}
\affiliation{Department of Physics and Astronomy, and LaserLaB, VU University, De Boelelaan 1081, 1081 HV Amsterdam, The Netherlands}
\author{J. Komasa}
\affiliation{Faculty of Chemistry, A. Mickiewicz University, Grunwaldzka 6, 60-780 Pozna\'n, Poland}
\author{K. Pachucki}
\affiliation{Faculty of Physics, University of Warsaw, Ho\.za 69, 00-681 Warsaw, Poland}
\author{K. S. E. Eikema}
\affiliation{Department of Physics and Astronomy, and LaserLaB, VU University, De Boelelaan 1081, 1081 HV Amsterdam, The Netherlands}
\author{W. Ubachs}
\affiliation{Department of Physics and Astronomy, and LaserLaB, VU University, De Boelelaan 1081, 1081 HV Amsterdam, The Netherlands}

\date{\today}

\begin{abstract}
Highly accurate results from frequency measurements on neutral hydrogen molecules H$_2$, HD and D$_2$ as well as the
HD$^+$ ion can be interpreted in terms of constraints on possible fifth-force interactions. Where the hydrogen atom is a probe for yet unknown lepton-hadron interactions, and the helium atom is sensitive for lepton-lepton interactions, molecules open the domain to search for additional long-range hadron-hadron forces.
First principles calculations in the framework of quantum electrodynamics have now advanced to the level that hydrogen molecules and hydrogen molecular ions have become \emph{calculable} systems, making them a search-ground for fifth forces.
Following a phenomenological treatment of unknown hadron-hadron interactions written in terms of a Yukawa potential of the form $V_5(r)=\beta \exp(-r/\lambda)/r$ current precision measurements on hydrogenic molecules yield a constraint
$\beta < 1 \times 10^{-7}$ eV$\cdot$\AA\ for long-range hadron-hadron interactions at typical force ranges commensurate with
separations of a chemical bond, \emph{i.e.} $\lambda \approx 1$  \AA\ and beyond.
\end{abstract}


\maketitle

\section{Introduction}

While the Standard Model (SM) of physics explains physical phenomena observed at the microscopic scale, phenomena of Dark Matter~\cite{Bertone2005} and Dark Energy~\cite{Peebles2003} at the cosmological scale are considered as unsolved problems, possibly hinting at physics beyond the SM. String theory~\cite{Aharony2000} and supersymmetry~\cite{Haber1985} seek to accomodate these phenomena, as well as gravity, with the SM within a unified model.
New kinds of fundamental interactions and/or extra dimensions are postulated as extensions of the SM~\cite{Hamed1998,Antoniadis1998}, which could be probed via high-energy particle colliders. However, there is also a frontier of low-energy physics~\cite{Jaeckel2010}, with predictions of weakly interacting particles at the eV energy scale, that could be probed in table-top experiments.

Celebrated examples of probing new physics at the atomic scale include experiments aimed at measuring an electric dipole moment of the electron, first through measurements on atoms~\cite{Regan2002}, and later at increased precision level also on molecules~\cite{Hudson2011}.
Experimental searches have been carried out to detect anomalous spin-spin interactions between electrons investigated through spectroscopy in ion traps~\cite{Wineland1991}, or in paramagnetic salts~\cite{Chui1993}. Similar anomalous spin-spin couplings between neutrons are investigated in $^3$He/$^{129}$Xe masers~\cite{Glenday2008}. Constraints on short-range spin-dependent interactions between protons and deuterons at the \AA\ scale are derived from nuclear magnetic resonance experiments in the HD molecule~\cite{Ledbetter2013}.

To date the energy level structure of atomic and molecular systems can be fully described by electromagnetism, in its most advanced form by the theory of quantum electrodynamics (QED). Effects of the weak force, leading to parity non-conservation, have been clearly observed at the atomic scale~\cite{Wood1997}, but not yet in light atoms as H/D and He. In these calculable systems effects of the weak interaction on the energy level structure is orders of magnitude away from experimental determination, although for muonic hydrogen the contribution is just below the accuracy of the proton-size contribution~\cite{Eides2012}.
Strong interactions, in quantum chromodynamics confined to the fm scale, are at the origin of nuclear $g_N$ factors and nuclear spins $I_N$, thus influencing atomic and molecular level energies in terms of hyperfine structure.
Obviously, gravitational interactions are far too weak to play any role in the calculation of level energies in atomic systems.
Hence the rationale for probing \emph{fifth forces} beyond the SM is based on a search for deviations from QED in the quantum level structure of \emph{calculable} systems at the atomic scale.

QED has been tested in light atoms to high precision, for example by the observed agreement in the derived proton rms charge radius $r_\mathrm{p}$ from several atomic hydrogen transitions.
In fact, by assuming the correctness of QED, the present CODATA recommendation~\cite{Mohr2012} for $r_\mathrm{p}$ is based on these derived values from high-precision hydrogen spectroscopy along with the value from electron-proton scattering experiments, where both derivations are in good agreement.
However, this $r_\mathrm{p}$ value is in 7$\sigma$-disagreement with that derived from recent measurements on muonic hydrogen ($\mu^-$p$^+$)~\cite{Pohl2010,Antognini2013}.
Although this deviation is interpreted as a puzzle on the proton size, the solution to this conundrum might as well be found in unaccounted effects within QED~\cite{Pohl2013}.

The He atom, a two-electron system, is still accessible for accurate \emph{ab initio} calculations of the level structure in the framework of QED~\cite{Yerokhin2010,Morton2006}, although measured level splittings are in several cases more accurate than theory~\cite{Kandula2010,Vanrooij2011,CancioPastor2012}. Precision measurements on the helium atom allow for tests of QED including lepton-lepton interactions that are not detectable in measurements of e$^-$H$^+$, nor in anti-hydrogen (e$^+$H$^-$), nor in muonic hydrogen.

Precision QED test have been extended to molecules, \emph{i.e.} to systems with more than one nucleon, where long-range hadron-hadron forces come into play. Quantum \emph{ab initio} calculations have been performed on the single-electron H$_2^+$ ionic system by solving the three-body Coulomb problem and calculating QED contributions up to high order in $\alpha$~\cite{Korobov2009,Karr2012}. Similar calculations of equal accuracy have been carried out for the HD$^+$ ion~\cite{Karr2005,Korobov2006,Korobov2012}, which exhibits a small dipole moment and is therefore more amenable for laser precision measurements~\cite{Koelemeij2007,Bressel2012}. In recent years, full-fledged level structure calculations, including QED and high-order relativistic contributions have also been carried out for the neutral hydrogen molecules and the deuterated isotopomers~\cite{Piszcziatowski2009,Komasa2011,Pachucki2010b}.
These calculations have been tested in the determination of dissociation limits of H$_2$~\cite{Liu2009}, D$_2$~\cite{Liu2010} and HD~\cite{Sprecher2010}, as well as in measurements of the ground state rotational sequence of \Hm~\cite{Salumbides2011}, and the vibrational splittings in \Hm, \Dm\ and HD \cite{Maddaloni2010,Kassi2012,Hu2012,Campargue2012,Dickenson2013}.

Fifth-force tests are commonly associated with testing non-Newtonian gravity, sometimes motivated by theories postulating extra dimensions~\cite{Hamed1998}. Tests of the inverse-square law behaviour of gravity have been carried out over a wide distance scale from kilometers to submicron, where the present short distance constraints are obtained from Casimir force experiments~\cite{Lamoreaux}.
We propose here that molecules open up a new arena for probing fifth forces at typical distance separations occurring in chemical bonds, thus at length scales naturally set to \AA\ distances. Attractive or repulsive additional forces can be probed from precision metrology measurements on calculable molecular systems. Based on recent precision measurements on HD$^+$ ions and H$_2$, D$_2$ and HD neutral molecules, and comparison with advanced QED calculations, constraints on a \emph{fifth-force} can be derived, which is parameterized by a generalized Yukawa potential for a certain effective range.
Where lepton-nucleon interactions and lepton-lepton interactions may be probed in atomic hydrogen and helium, in the present study additional forces between hadrons at long range are targeted, for which molecules are a good test ground.

\section{QED calculations in neutral and ionic molecular hydrogen}

Accurate \emph{ab initio} level energies $E$ of molecular hydrogen and its deuterated isotopomers are calculated in the framework of nonrelativistic quantum electrodynamics (NRQED) from an evaluation in orders of the electromagnetic coupling constant $\alpha$
\begin{equation}
E(\alpha) = \,{\cal E}^{(0)} + \alpha^2\,{\cal E}^{(2)} + \alpha^3\,{\cal E}^{(3)} + \alpha^4\,{\cal E}^{(4)} + ...
\label{E}
\end{equation}
The nonrelativistic energy ${\cal E}^{(0)}$ is obtained by solving the Schr\"odinger equation using the Born-Oppenheimer potential with 15 digit accuracy~\cite{Pachucki2010a}. Adiabatic and nonadiabatic corrections are subsequently calculated perturbatively in powers of the electron-nucleus mass ratio.
This procedure results in nonrelativistic binding energies accurate to a few parts in $10^{-4}$ cm$^{-1}$~\cite{Pachucki2009}.
The $u-g$ symmetry-breaking is taken into account for the specific case of HD\cite{Pachucki2010b}.
Leading-order relativistic corrections ${\cal E}^{(2)}$ have been calculated from the expectation value of
the Pauli Hamiltonian~\cite{Piszcziatowski2009,Komasa2011}.
The leading QED corrections of order ${\cal E}^{(3)}$ are treated similarly as for the hydrogen and helium atoms~\cite{Drake2000}.
This results in an accuracy of $10^{-6}$ cm$^{-1}$ for ${\cal E}^{(3)}$ in molecular hydrogen due to neglect of nonadiabatic and relativistic recoil corrections.
The main contribution to the calculation uncertainty at present is from the ${\cal E}^{(4)}$ QED correction,
which has not been calculated explicitly due to the high complexity of NRQED operators.
The radial nuclear functions for $v=0$ and $v=1$ probe almost the same range of internuclear distance, leading to significant cancellation of the uncertainty of the energy contributions.
The final theoretical predictions are estimated to be accurate to $1 \times 10^{-4}$ cm$^{-1}$ for the full QED evaluation of the rotationless fundamental ground tones in H$_2$, D$_2$ and HD. There is less cancellation in uncertainty of transition energies as the difference in the vibrational quantum number $\Delta v$ increases, leading to larger uncertainties at $\sim 1 \times 10^{-3}$ cm$^{-1}$ for the dissociation energies ($D_0$) of molecular hydrogen and its isotopomers.

A reduction in complexity of three-body system molecular ions compared to the (two-electron) neutral molecular species results in better calculation accuracy for \Hmi\ and \HDi. Nonrelativistic energies are obtained with up to 30-digit numerical precision for the low-lying vibrational states~\cite{Li2007} using a direct variational approach. Relativistic and QED corrections are also calculated using the NRQED framework similarly expressed as Eq.~(\ref{E}) for the neutral species. The leading-order relativistic corrections ${\cal E}^{(2)}$ have been calculated with sub-kHz accuracy~\cite{Zhong2009}.
A recent improvement in the evaluation of the Bethe logarithm~\cite{Korobov2012} has enabled an increased accuracy of the ${\cal E}^{(3)}$ term at better than 50 Hz. The total uncertainty is dominated by the contribution of the higher order QED terms ${\cal E}^{(4)}$ and ${\cal E}^{(5)}$, estimated to be $\sim20$ kHz.
Note that in the case of the ion with its unpaired electron, the hyperfine interaction (which is absent to first order in the neutrals) is addressed separately~\cite{Bakalov2006}. The QED calculation uncertainty contributes 21 kHz while the (in)accuracy of fundamental constants contributes 10 kHz to the total uncertainty of the \emph{ab initio} calculation for the \HDi\ $v=0\rightarrow 1$ R(1) transition, corrected for hyperfine structure~\cite{Bressel2012}.

\section{Precision measurements in molecules}
\label{Experimental}

High-precision molecular spectroscopy on neutral and ionic molecular hydrogen is reviewed in this section. The excellent agreement found between these experimental results with \emph{ab initio} calculations, provide the most stringent tests on the application of quantum electrodynamics in a chemically-bound system.

Recently, the rotationless vibrational transitions ($v''=0\rightarrow v'=1$) for \Hm, HD, and \Dm\ were determined to an accuracy of $\sim 2 \times 10^{-4}$ cm$^{-1}$ \cite{Dickenson2013}.
These fundamental ground tone vibrations of H$_2$, HD, and D$_2$ were obtained from Doppler-free laser spectroscopy in the collisionless environment of a molecular beam.
The rotationless fundamental vibrational splitting was derived from the combination difference between electronic excitation
from the $X^1\Sigma_g^+, v=0$ and $v=1$ levels to a common $EF^1\Sigma_g^+, v=0$ level.
The experimental results are in excellent agreement with a full \emph{ab initio} calculation up to an uncertainty $\delta E\sim 2\times10^{-4}$ \wn, where the combined precision of the experimental and theoretical values is defined as $\delta E = \sqrt{\delta E_\mathrm{exp}^2 + \delta E_\mathrm{calc}^2}$.

Some overtone transition frequencies of molecular hydrogen have been determined via single-photon infrared absorptions, with the most recent investigations employing cavity-ringdown (CRDS) techniques \cite{Hu2012,Campargue2012,Kassi2012}. Despite the sensitivity of CRDS, the extremely weak quadrupole transitions necessitate high-pressure samples subject to pressure shifts in addition to Doppler-broadening. The results from Hu \et~\cite{Hu2012}, Campargue \et~\cite{Campargue2012} and Kassi \et~\cite{Kassi2012} are in agreement with theory, and result in a combined precision of the comparison of $\delta E\sim 1\times10^{-3}$ \wn.
Maddaloni~\et~\cite{Maddaloni2010} claim an absolute accuracy for the S(0) and S(2) transitions of the fundamental ground tone in D$_2$ at $2\times10^{-4}$ cm$^{-1}$, however, a comparison to the theoretical value yields a $7\sigma$ discrepancy.
Owing to this large inconsistency from Ref.~\cite{Maddaloni2010}, bounds derived from these transitions are not included.

The recent and most accurate experimental determinations on the dissociation limits of H$_2$~\cite{Liu2009}, D$_2$~\cite{Liu2010} and HD~\cite{Sprecher2010}, are based on three energy intervals obtained from separate spectroscopic investigations. The ionization potential (IP) is the sum of these energy intervals: the first interval is between the ground electronic state and the EF $v=0$, $J$ state; the second between is the EF state to a high-$n$p Rydberg state; and the third is between the high-$n$p state to the molecular ion H$_2^+$ ground state X $v=0$. The neutral molecule dissociation limit is derived by combining the IP with the accurate theoretical calculation of the molecular ion dissociation energy and the accurate experimental value of the atomic ionization energy. These neutral molecule dissociation limits are in excellent agreement with the most accurate \emph{ab initio} calculations, again demonstrating the correctness of QED evaluations at $\delta E\sim 1\times10^{-3}$ \wn\ level.

High-resolution molecular spectroscopy has been performed on trapped and cooled \HDi~\cite{Koelemeij2007,Bressel2012} as well.
The use of sympathetic cooling of \HDi\ by optically cooled Be$^{+}$ ions reduces the Doppler widths of the molecular transitions.
In a resonance-enhanced multiphoton dissociation (REMPD) scheme, the ground state \HDi\ ion is first optically excited to a higher vibrational quantum state.
Thereafter, a second photon further excites the ion to a dissociative state leading to a loss of the trapped \HDi\ which can be detected.
Excellent agreement between experiment and theory is observed for the spectroscopic results on the $v''=0\rightarrow v'=4$ transition~\cite{Koelemeij2007} in \HDi, with the experimental accuracy of the hyperfineless transition at $1.7\times10^{-5}$ \wn. The theoretical value is accurate to $2.3\times10^{-6}$ \wn\ resulting in a combined uncertainty of $\delta E=1.7\times10^{-5}$ \wn.

The hyperfine structure is partially-resolved in a recent vibrational spectroscopy investigation of the $v=0\rightarrow1$ band in \HDi~\cite{Bressel2012} leading to an improved measurement accuracy.
However, there is currently a $2\sigma$ discrepancy between theory and experiment for the hyperfineless transition energy for this fundamental vibrational splitting.
It appears plausible that this deviation is caused by statistical noise or an as of yet unaccounted hyperfine interaction.
Therefore, rather than considering this a possible manifestation of a fifth force, we include this transition in the fifth force constraining analysis with $\delta E$ set equal to the $2\sigma$ discrepancy between theory and experiment.

Experiments on the $v=0\rightarrow8$ band in \HDi\ are in preparation, with great potential for probing fifth force interactions, but no precision frequency results have been obtained yet~\cite{Koelemeij2012}.

The most accurate dataset that is used to further derive constraints for \emph{fifth-force} interactions is listed in Table~\ref{data}.

\begin{table}
\caption{Relevant data from neutral and ionic molecular hydrogen transitions used to derive constraints.
}
\label{data}
\begin{tabular}{l@{\hspace{15pt}}c@{\hspace{15pt}}r@{\hspace{15pt}}c}
\toprule
species & transition & $\delta E$ (\wn) & ref. \\
\colrule
\Hm	&$v=0\rightarrow1$ 	&0.00020	&\cite{Dickenson2013} \\
	&$v=0\rightarrow2$	&0.004		&\cite{Campargue2012} \\
	&$v=0\rightarrow3$	&0.004		&\cite{Hu2012} \\
	&$D_0$			&0.0012		&\cite{Liu2009} \\
\colrule
HD	&$v=0\rightarrow1$ 	&0.00025	&\cite{Dickenson2013} \\
	&$D_0$			&0.0012		&\cite{Sprecher2010} \\
\colrule
\Dm	&$v=0\rightarrow1$ 	&0.00018	&\cite{Dickenson2013} \\
	&$v=0\rightarrow2$	&0.001		&\cite{Kassi2012} \\
	&$D_0$			&0.0011		&\cite{Liu2010} \\
\colrule
\HDi	&$v=0\rightarrow1$ 	&0.000005\footnote{$\delta E$ is the discrepancy between theory and experiment} &\cite{Bressel2012} \\
	&$v=0\rightarrow4$ 	&0.000017	&\cite{Koelemeij2007} \\

\botrule
\end{tabular}
\end{table}

\section{Search for fifth forces in molecules}

The occurrence of fifth forces beyond the SM can be phenomenologically parameterized by a Yukawa-type potential with an effective range $\lambda$
\begin{eqnarray}
	V_5(r) = \beta^\prime \frac{\exp{(-r/\lambda})}{r} = \beta^\prime Y(r)
\label{Y0}
\end{eqnarray}
where $\beta^\prime$ is a coupling strength, which may \emph{a priori} be attractive or repulsive.
An observation of a significant effect would prompt a more thorough investigation of particular model potentials that best represent an observed discrepancy.
The presence of a Yukawa potential correction implies the existence of a force carrier with a mass inversely proportional to $\lambda$. For the effective separation distance on the \AA-scale relevant to this study, this may be viewed as searches for effects of new force-carrier particles with light masses in the order of keV, where the particle interaction length is taken to be the Compton wavelength ($h/mc$) of a hypothetical bosonic gauge particle.

In this particular case of molecules, the effect of a \textit{fifth-force} between nuclei can be searched for at the distances where they are bound within the geometry of the molecules; in this case $r$ is the internuclear distance. To incorporate the effects of different nucleon numbers $N$ in H$_2$, HD, and  D$_2$, as well as the H$_2^+$ and HD$^+$ ions a redefinition of the coupling constant is introduced to explicitly express the dependence:
\begin{equation}
	V_5(r) = \beta N_{1} N_{2} Y(r),
\label{Y1}
\end{equation}
where $N_{1,2}$ are the nucleon numbers for each nucleus; note that this differs from a definition by Bordag~\emph{et al.}~\cite{Bordag1994}. The extra long-range hadron-hadron interaction probed here is spin-independent. In view of the nucleon scaling factors, molecules with the highest number of nucleons will provide the tightest constraint on the existence of fifth forces parameterized by $V_5(r)$; hence the D$_2$ isotopomer would be a more sensitive test ground than H$_2$.

We treat the extra potential in Eq.~(\ref{Y1}) as a perturbation on the ro-vibrational level energies of the molecular states. For a transition between the ground $v''$ and excited $v'$ vibrational levels, the contribution $\Braket{\Delta V_5}$ of a fifth-force potential can be expressed as
\begin{eqnarray}
	\Braket{ \Delta V_{5,\lambda} }  &=&  \beta N_{1} N_{2} \left[ \Braket{ \Psi_{v',J'}(r)  |  Y(r,\lambda) | \Psi_{v',J'}(r) } \right. \nonumber \\
                   && \qquad \qquad \left. - \Braket{ \Psi_{v'',J''}(r)  |  Y(r,\lambda) | \Psi_{v'',J''}(r) }  \right] \nonumber \\
					&=& \beta N_{1} N_{2} \Delta Y_{\lambda}
\label{YV5}
\end{eqnarray}
where $\Psi_{v,J}(r)$ represents the nuclear wave functions. Although included in the calculations, the explicit mention of rotational quantum numbers $J$ is omitted in the rest of the discussion since rotation has a much smaller contribution compared to vibration for the transitions treated here.

\begin{figure}
\includegraphics[width=\columnwidth]{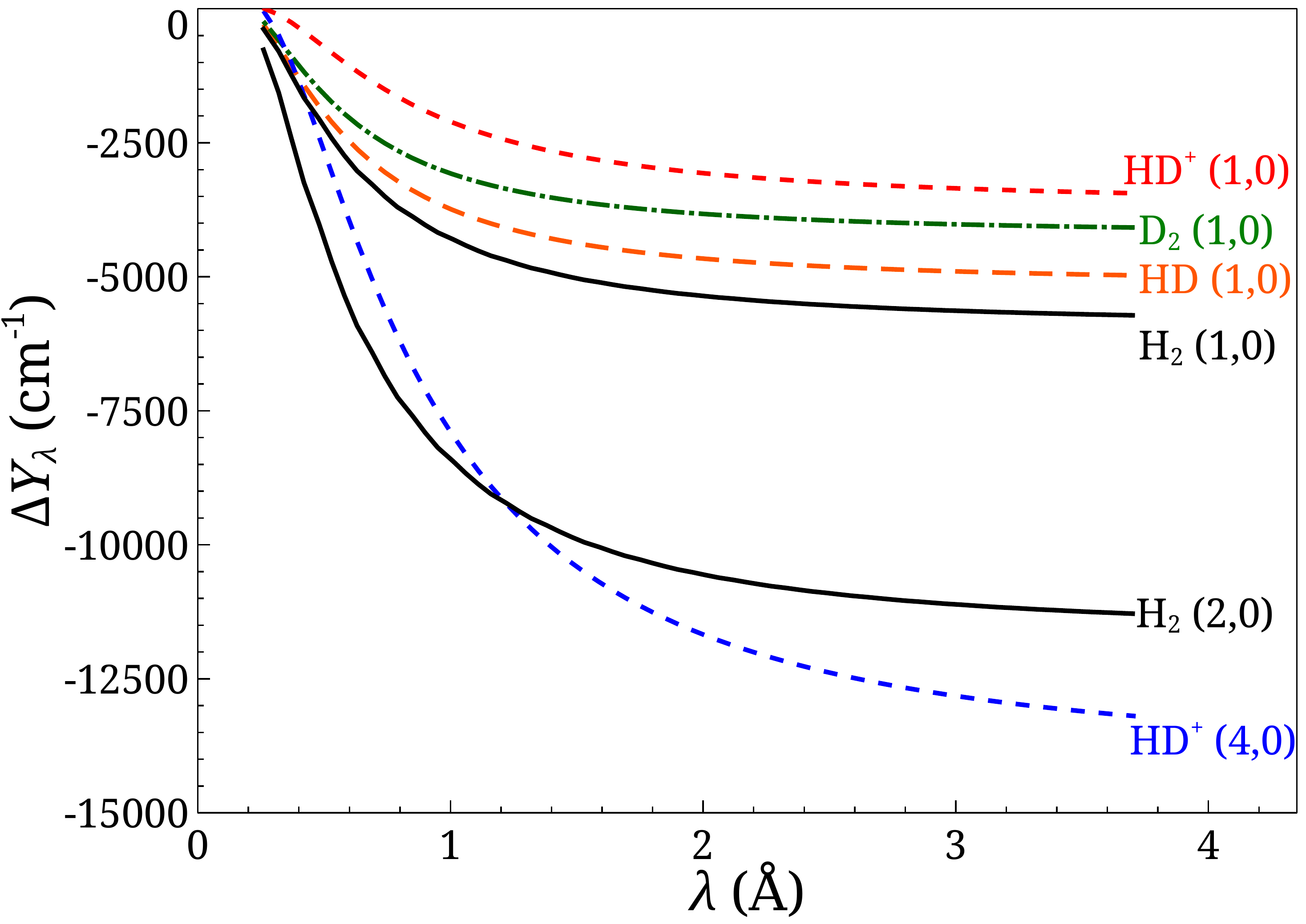}
\caption{
(Color online) Calculated difference of the expectation values $\Delta Y_{\lambda}$ for different values of $\lambda$ for several ($v',v''=0$) vibrational energy separations in \Hm, \Dm, HD and \HDi\ molecules.
\label{Fig1}}
\end{figure}

Numerical calculations were performed to evaluate the $\Braket{\Delta V_{5,\lambda}}$ contribution of various electronic ground state vibrational transitions using the accurate wave functions for \Hm, HD, and \Dm\ \cite{Piszcziatowski2009,Komasa2011,Pachucki2010b} and \HDi\ \cite{Koelemeij2011}.
The difference of the expectation values $\Delta Y_{\lambda}$ in Eq.~(\ref{YV5}) is plotted in Fig.~\ref{Fig1} for the $v''=0\rightarrow v'=1,2$ transitions in \Hm, with the interaction length $\lambda$ taken as a parameter. The corresponding differential contribution for vibrational transitions for $v''=0\rightarrow v'=1$  for \Dm\ and HD and for $v''=0\rightarrow v'=1,4$ in \HDi\ are also plotted in Fig.~\ref{Fig1}.

This illustrates that for a specific molecule, the sensitivity for probing a fifth-force contribution increases as $\Delta v$ increases. This can be attributed to the difference in the spatial extent of the ground and excited state wave functions, where the latter has a most probable position (related to classical bond length) gradually displaced from the equilibrium for increasing $v$ quantum numbers. Since the wave function density for consecutive vibrational levels shifts only slightly, of effect of a Yukawa potential $V_5$ is probed in molecules as a differential contribution.
For a given vibrational transition, \Hm\ has the greater sensitivity compared to \Dm\ and HD (before the nucleon-number scaling) since the wave functions belonging to different vibrational levels have more similar spatial extents in the heavier isotopes than in \Hm.
For specific vibrational transitions, e.g.  ($v''=0\rightarrow v'=1$), the differential contribution for \Hm\ is also greater than the corresponding contribution \HDi\ since the distance between the nuclei in the ions, with only one electron contributing to the chemical bond, is larger. In general, transitions in molecules with shorter bond lengths will be more sensitive to a fifth-force hadron-hadron potential.

Any difference $\delta E$ between the experimental and calculated values for a particular transition energy can be used to set bounds for the maximum contribution to a fifth-force. A constraint for the coupling strength $\beta$ is obtained for a range of values of an interaction length $\lambda$ by the relation
\begin{equation}
\beta < \frac{\delta E}{N_{1} N_{2}\Delta Y_{\lambda}}.
\end{equation}

The coupling constant $\beta$ has units of energy$\times$distance, and can also be related to any known interaction such as the electromagnetic interaction characterized by the coupling constant $\alpha$.
The Coulomb potential can be expressed as $V_{em}=\alpha Z_1 Z_2/r$ with units of energy/distance, where $Z_{1,2} = 1$ for the nuclei considered in this study. Except for the extra exponential factor, $V_5$ has the same $1/r$-form and hence the same dimensions as $V_{em}$, leading to a dimensionless ratio of their interaction strengths $\beta/\alpha$.     

\subsection{The fundamental vibration in the hydrogen molecule}

The agreement of the accurate experimental and theoretical values are used to provide a constraint on the effect of new interactions. The combined experimental and theoretical uncertainty at $\delta E = 2 \times 10^{-4}$ \wn\ for the fundamental vibrational tone provides limits on the interaction strength $|\beta|$ for different values of the interaction length $\lambda$.
The limits derived from the \Hm\ $X,v''=0\rightarrow v'=1$ transition constrain the strength of a new interaction to be $\beta < 4.7 \times 10^{-8} \alpha$ for interaction lengths $\lambda > 1$ \AA. Similarly bounds are obtained using the HD and \Dm\ fundamental vibrational tone at $\beta < 3.4 \times 10^{-8} \alpha$ and $\beta < 1.5 \times 10^{-8} \alpha$, respectively, for interaction lengths $\lambda > 1$ \AA.

\subsection{Overtone vibrations in neutral hydrogen}

\begin{figure}
\includegraphics[width=\columnwidth]{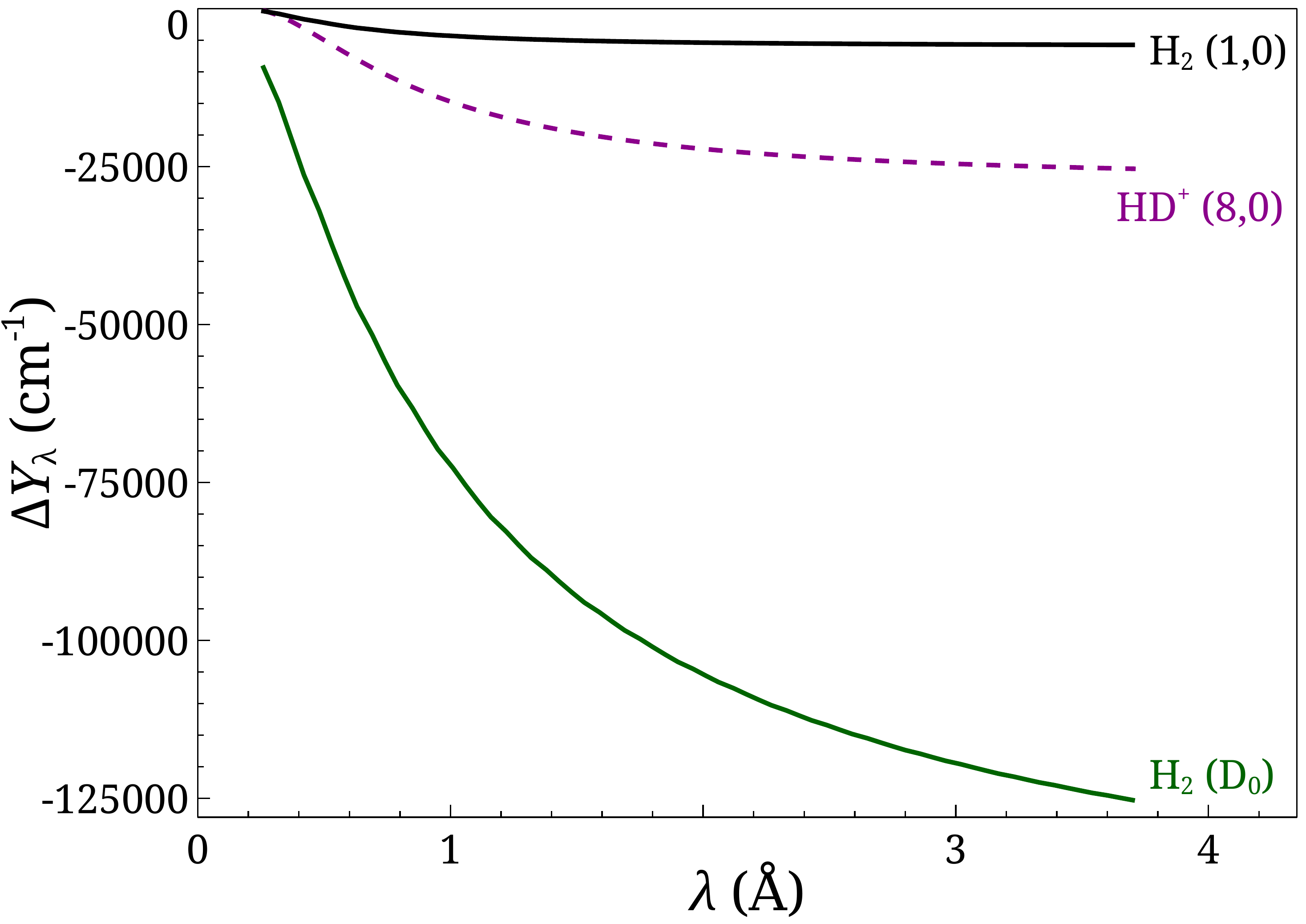}
\caption{
(Color online) Calculated expectation value $Y_{\lambda,D_0}$ of a fifth-force contribution to the \Hm\ dissociation limit $D_0$ for a range of $\lambda$ values. For reference the contributions $\Delta Y_{\lambda}$ for the ($v'=1,v''=0$) band of \Hm\ and the ($v'=8,v''=0$) band of HD$^{+}$ electronic ground state are plotted as well.
\label{expectation}}
\end{figure}

Similar constraints can be derived from investigations of the electronic ground state direct overtone vibrations (e.g. $v^\prime=0$ to $v=2,3$) of \Hm\ and \Dm.
The results from recent CRDS studies on the overtone quadrupole transitions of \Hm\ \cite{Campargue2012,Hu2012} and \Dm\ \cite{Kassi2012,Maddaloni2010} can be used to extract constraints. The overtone transitions of \Hm\ and \Dm\ have intrinsically \emph{higher} sensitivity of the transitions in comparison to the fundamental ground tone as shown in Fig.~\ref{expectation}. However, the worse uncertainty ($\sim10^{-3}$ cm$^{-1}$) from these Doppler-limited studies, and pressure-shift corrected transition energies, result in less tight constraints. For interaction lengths $\lambda > 1$ \AA, the \Hm\ (2,0) band constrains $\beta < 4.7 \times 10^{-7} \alpha$ and the (3,0) band constrains $\beta < 2.4\times10^{-7} \alpha$, while the \Dm\ (2,0) band leads to a constraint of $\beta < 4.2 \times 10^{-8} \alpha$.

\subsection{Level energies in the HD$^+$ molecular ion}

The spectroscopic results for \HDi\ for the $v''=0\rightarrow v'=4$ transition~\cite{Koelemeij2007} provide stringent bounds on possible fifth-force interactions.
For $\lambda > 1$ \AA, the  constraint for the interaction strength is $\beta < 1.1 \times10^{-9} \alpha$. This tighter bound from \HDi\ is due to the better accuracy in both experiment and theory, as well as larger $\Delta v$ probed between the $v=0$ and $v=4$ quantum states.

The results for \HDi\ vibrational spectroscopy of the $v=0\rightarrow1$ transitions can still be used to also provide bounds for new hadronic interactions.
The two-sigma discrepancy between theory and experiment $\delta E = 4.7 \times 10^{-6}$ \wn\ is used to derive bounds to a fifth-force interaction
with $\beta < 1.1\times10^{-9} \alpha$ for $\lambda > 1$ \AA.

\subsection{The dissociation limit in H$_2$, HD and D$_2$}

The dissociation energy is defined to be the energy difference between the deepest bound molecular state $v''=0, J''=0$ and the state when the two constituent atoms are non-interacting, i.e. at $r=\infty$ where $V(\infty)=0$. In a similar way, a possible fifth-force interaction is treated perturbatively and the contribution is expressed as
\begin{eqnarray}
	\Braket{\Delta V_{5,\lambda}}  &=& - \beta N_1 N_2 \Braket{\Psi_{v''=0}(r)  | Y(r,\lambda) | \Psi_{v''=0}(r)} \nonumber \\
				&=& - \beta N_1 N_2 Y_{\lambda,D_0}.
\label{Ydiss}
\end{eqnarray}
where $\lambda$ is again treated as a parameter in the calculations.

The fifth force contribution to the dissociation limit is the upper limit of the $\Delta v$-progression for the vibrational transitions represented in Eq.~(\ref{YV5}), and thus intrinsically is more sensitive than any of the ground state vibrational splitting.

The combined experiment--theory accuracy of $D_0$ for H$_2$ is $\delta E = 1.2 \times 10^{-3}$ \wn~Ref.~\cite{Liu2009}; for D$_2$~\cite{Liu2010} it is $\delta E =1.1 \times 10^{-3}$ \wn; and for HD~\cite{Sprecher2010} it is $\delta E=1.1 \times 10^{-3}$ \wn. The \Dm\ value gives the tightest constraint at $\beta < 3.8 \times 10^{-9} \alpha$ for interaction lengths $\lambda > 1$ \AA, \Hm\ and HD gives $\beta < 1.7 \times 10^{-8} \alpha$ and $\beta < 7.5 \times 10^{-9} \alpha$, respectively, for $\lambda > 1$ \AA.

\begin{figure}
\includegraphics[width=\columnwidth]{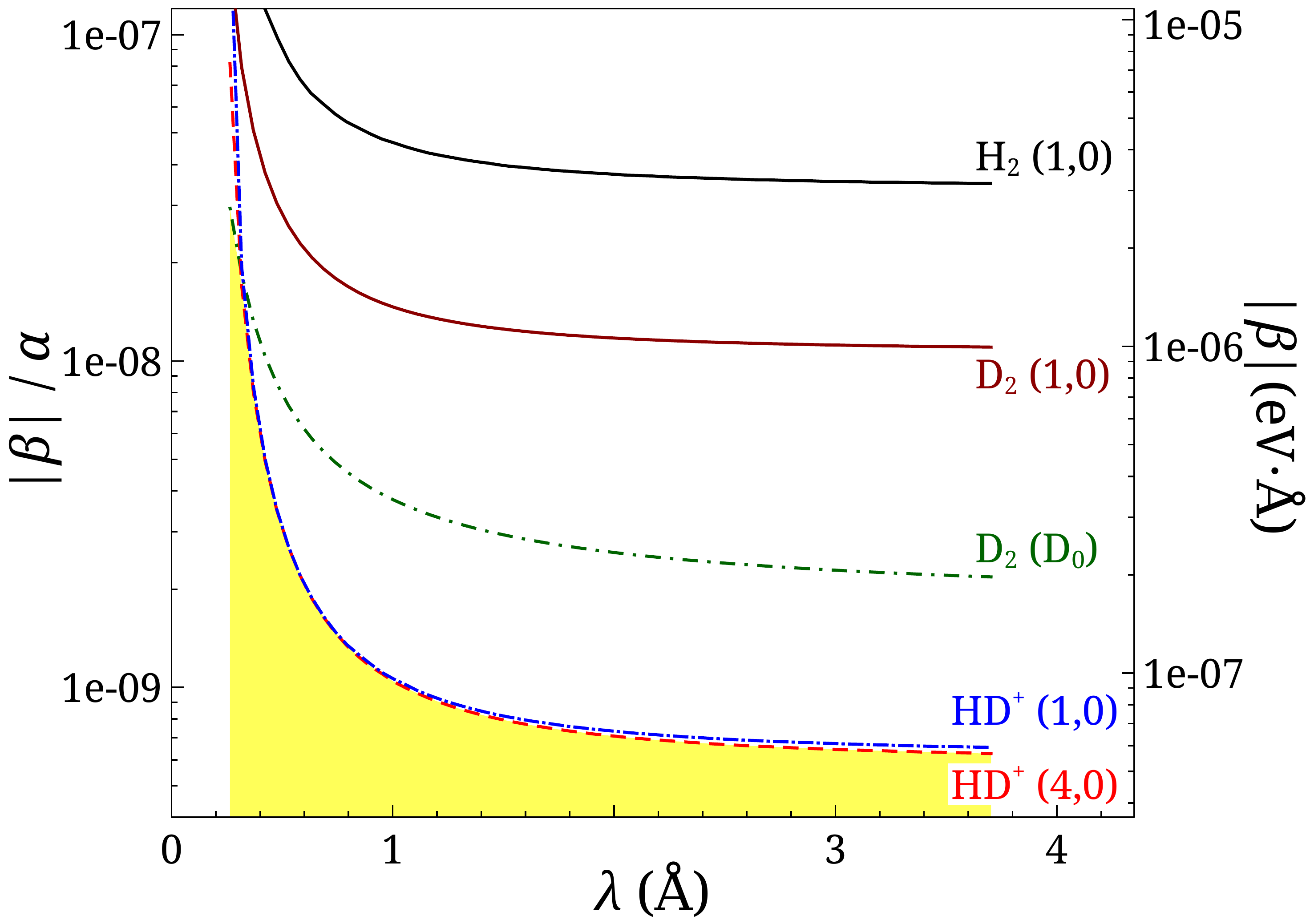}
\caption{
(Color online) Limits on the strength of the coupling constant $\beta$ relative to the electromagnetic coupling constant $\alpha$ as a function of the interaction range $\lambda$. The alternate axis expresses $\beta$ in units of eV$\cdot$\AA.
\label{limits}}
\end{figure}

\section{Discussion}

Constraints on a fifth-force coupling constant $\beta$, for a phenomenological Yukawa interaction potential $V_5(r)=\beta \exp(-r/\lambda)/r$, are derived from vibrational transitions in both neutral and ionic molecular hydrogen species measured at high precision. Some of the tightest constraints are plotted in Fig.~\ref{limits}, where the value of $|\beta| / \alpha$ is used to express the strength of a fifth-force with respect to the strength $\alpha$ of the electromagnetic interaction, with an alternative axis expressing $\beta$ in units of eV$\cdot$\AA.
The region below a curve represents the allowed value of $\beta$ for a certain value of the range of the force $\lambda$. The constraints obtained from \HDi\ are most stringent owing to the better accuracy in both the experimental and calculated values. At present, the \HDi\ (4,0) band furnishes slightly tighter limits than that of the (1,0) band, both because of the inherent enhancement ($\Delta Y_{\lambda}$) of the former and also the 2$\sigma$ deviation between theory and experiment in the latterwhich might be due to a statistical error or to an as of yet unaccounted hyperfine interaction. The tighter constraint from \Dm\ compared to \Hm, gained from the nucleon number scaling, is demonstrated graphically in Fig.~\ref{limits}. The present analysis yields a consistent constraint $\beta/\alpha < 10^{-9}$ or $\beta < 1 \times 10^{-7}$ eV$\cdot$\AA\ for long-range hadron-hadron interactions at $\lambda > 1$  \AA.
The allowed value for $\beta$ is indicated by the shaded region in Fig.~\ref{limits}.

The present bounds obtained from the \Dm\ dissociation limit is only a factor four less tight compared to that of the \HDi\ (4,0) band, despite the experimental uncertainty of \Dm\ $D_0$ being $\sim$60 times worse than for the ion. This suggests that future improvements in the accuracy of the \Dm\ $D_0$~\cite{Sprecher2011} can lead to tighter constraints, while in the case of \HDi\, hyperfine interaction need to be addressed more accurately to improve its accuracy. In addition, as the wave functions of the neutral hydrogen molecules extend towards shorter range than the ionic species, the bounds for $\beta$ at shorter $\lambda$ is tighter for the neutrals, \emph{e.g.} for $\lambda < 0.4$ \AA\ in Fig.~\ref{limits} $D_0$ of \Dm\ gives the tightest constraints.

Precision measurements on exotic molecules, such as anti-protonic helium~\cite{Hori2011}, could provide similar constraints for force ranges on shorter length scales, \emph{i.e.} $\lambda < 0.5$ \AA.

\section{Conclusion}

The advancements in recent years on the \emph{ab initio} theory of light molecular species, e.g. neutral and ionic molecular hydrogen and its isotopomers, have led to the successful application of QED corrections in quantum chemical calculations. Comparison of these theoretical results with highly accurate experimental values from precision spectroscopic investigations have demonstrated excellent agreement between theory and experiment.
Treating fifth force long-range hadron-hadron interactions in the form of a Yukawa-type interaction potential, we show how such highly accurate comparisons between theory and experiment provide a search ground for new physics beyond the Standard Model.

\section*{Acknowledgments}
This research was supported by the Netherlands Foundation for Fundamental Research of Matter (FOM) through the program "Broken Mirrors \& Drifting Constants".
W.U. thanks the Templeton Foundation for a New Frontiers in Cosmology grant. J.C.J.K. thanks the Dutch Organisation for Scientific Research (NWO) for support.
Support by the NCN grants N-N204-015338 (J.K.) and 2012/04/A/ST2/00105 (K.P.) is acknowledged as well as by a computing grant from Poznan Supercomputing and
Networking Center, and by PL-Grid Infrastructure.

\end{document}